\documentstyle[preprint,aps]{revtex}

\tightenlines

\begin{document}

\draft

\title{\bf Various series expansions for \\
a Heisenberg antiferromagnet model for SrCu$_2$(BO$_3$)$_2$}
\author{Zheng Weihong\cite{byline1}, 
C.J. Hamer\cite{byline2} and J. Oitmaa\cite{byline3}
} 
\address{School of Physics,                                              
The University of New South Wales,                                   
Sydney, NSW 2052, Australia.}                      

\date{Nov. 3, 1998}

\maketitle 

\begin{abstract}
We use a variety of series expansion methods at both zero and finite 
temperature to study an antiferromagnetic Heisenberg spin model
proposed recently by Miyahara and Ueda for the quasi two-dimensional
material SrCu$_2$(BO$_3$)$_2$. We confirm that this model exhibits a first-order
quantum phase transition at $T=0$ between a gapped dimer phase and 
a gapless N\'eel phase when the ratio $x=J'/J$ of nearest and
next-nearest neighbour interactions is varied, and locate the
transition at $x_c=0.691(6)$. Using longer series we are able
to give more accurate estimates of the model parameters
by fitting to the high temperature susceptibility data.
\end{abstract}                                                              
\pacs{PACS numbers:  75.10.Jm., 75.40.Gb  }


\narrowtext
\section{INTRODUCTION}
The discovery of high-temperature superconductivity
has stimulated an enormous amount of activity in the
study of two-dimensional antiferromagnetism, which may be connected
with the superconductivity phenomenon. The pseudo spin-gap behaviour 
observed in the high T$_c$ cuprates has also stimulated intense
interest in systems with spin gaps. Several new spin gap systems have 
been found experimentally. Among them
 some of the compounds
which have two dimensional character include the coupled spin
ladder systems, SrCu$_2$O$_3$\cite{srcu2o3}, CaV$_2$O$_5$\cite{cav2o5},
(VO$_2$)P$_2$O$_7$\cite{vopo}, Cu$_2$(C$_5$H$_{12}$N$_2$)$_2$Cl$_4$\cite{cuHpCl},
and the plaquette RVB system, CaV$_4$O$_9$\cite{cav4o9}.

Recently a new two dimensional spin gap system SrCu$_2$(BO$_3$)$_2$ has been found
by Kageyama {\it et al.}\cite{kag98}. It has a spin-singlet ground
state with a finite spin gap $\sim 20$ K. They also found that the peak of
the susceptibility is much suppressed compared with standard dimer models,
and observed two plateaus in the magnetization at
1/4 and 1/8 of the full moment.

Miyahara and Ueda\cite{miy98} showed that these observations could be 
understood on the basis of a simple two-dimensional Heisenberg
antiferromagnet model with nearest-neighbor and next-nearest-neighbor
couplings. The copper ions in the SrCu$_2$(BO$_3$)$_2$ compound are all located at crystallographically
equivalent sites, forming a distinctive pattern. 
Miyahara and Ueda\cite{miy98} show that, remarkably enough, a singlet
dimer state forms an exact eigenstate of this Hamiltonian at all couplings, and
is the ground-state in a region where the nearest-neighbour
coupling dominates. There is a first order phase transition to
a N\'eel ordered state at $J'/J=0.7\pm 0.01$. They find that the
SrCu$_2$(BO$_3$)$_2$ system lies close to this transition, which explains the
unusual temperature dependence of the magnetization.
The plateaus observed in the magnetization curve can also be understood,
on the basis that the triplet excitations from the ground state are almost
localized. Their conclusions were reached on the basis of exact
diagonalization calculations for lattices of up to
20 sites, and low order dimer expansions in $(J'/J)$,
and high temperature expansions in $J/T$ and $J'/T$.

In this paper, we reinforce these conclusions by carrying out a more
extensive series study of the model. These include high-temperature
expansions, Ising expansions at zero temperature, and dimer
expansions both at zero and finite temperature.
The model is presented in Section II, and the framework
for the various series expansions is outlined.
The results are presented in Section III. It is shown that
the temperature dependence of the susceptibility can be fitted accurately
and in detail by this method.

\section{Series Expansions}

The magnetic properties of SrCu$_2$(BO$_3$)$_2$  may be described by the
two-dimensional spin $S=1/2$ Heisenberg antiferromagnetic model
with nearest-neighbor (n.n.) and next-nearest-neighbor (n.n.n.)
interactions\cite{kag98,miy98}:
\begin{equation}
H = J \sum_{{\rm n.n.}} {\bf S}_i\cdot {\bf S}_j 
+ J' \sum_{{\rm n.n.n.}} {\bf S}_i\cdot {\bf S}_j~. \label{H}
\end{equation}
The system is illustrated in Fig. 1(a). 
We denote the ratio of the couplings as $x$,
that is, $x \equiv J'/J$.
In the present paper, we study only the case of antiferromagnetic coupling, 
where both $J$ and $J'$ are positive.
In the large $J'/J$ limit,
the model is topologically equivalent
to the two dimensional nearest-neighbor square lattice Heisenberg model.
In the small $J'/J$ limit, every pair of spins 
along nearest-neighbor bonds interact only weakly
with each other, and the dominant configuration 
in the ground state is the product state with the spins 
along nearest-neighbor bonds forming a spin singlet.
In fact, 
it has been proved\cite{shastry,miy98} that this state is an eigenstate of
the system for any $x$,
and it is the ground state for small enough
values of $x$. We can also prove that
this system still has this perfect dimer state as an eigenstate
even if we include the coupling $J_2$ denoted by dotted lines 
shown in Fig. 1(b).

We have studied this system by using various linked-cluster expansion methods including
dimer expansions at both zero-temperature and finite-temperature, 
Ising expansions at zero-temperature, and
high temperature expansions.
The linked-cluster expansion method
has been previously reviewed in several
articles\cite{he90,gel90,gelmk,otja92,rajiv98}, and will not be repeated here.
Here we only summarize
the expansion methods used, and the results derived from them 
are presented in the next Section.

\subsection{Dimer Expansions at $T=0$}
At temperature $T=0$, we can construct an expansion in $x$ by taking
the first term of $H$ as the unperturbed Hamiltonian and the second term
in $H$ as a perturbation. That is, the Hamiltonian of Eq.(\ref{H}) can
be rewritten as
\begin{equation}
H = H_0 + x V~, \label{Hdimer}
\end{equation}
where 
\begin{eqnarray}
H_0 &=& \sum_{{\rm n.n.}}  {\bf S}_i\cdot {\bf S}_j ~, \nonumber \\ & & \\
V &=& \sum_{{\rm n.n.n.}} {\bf S}_i\cdot {\bf S}_j ~, \nonumber
\end{eqnarray}
and where we have set $J=1$ for convenience.
The unperturbed ground state is then a product state of nearest neighbor
singlet dimers and the perturbation couples these among themselves and
with the pair triplet states. As mentioned above, the unperturbed ground
state is also an eigenstate of the full Hamiltonian, but is not the
true ground state for $x>x_c$.

Dimer expansions can be developed for all ground state properties as well
as for the triplet excitation spectrum. Here, because of the trivial
nature of the ground state in the dimer phase, we concentrate on the
lowest triplet excitations. We have calculated the dispersion relation
$\Delta (k_x, k_y)$ to order $x^{15}$, extending the calculation of
Miyahara and Ueda by 11 terms. This
calculation involves 11586 linked clusters up to 8 sites.
The resulting series coefficients are given in Table I.

\subsection{Ising Expansions at $T=0$}
In the limit that $J'\gg J$, the model is topologically equivalent
to the two dimensional square lattice Heisenberg model,
so we expect that the system has Ne\'{e}l order:
an order in which every pair of 
spins along a horizontal nearest-neighbor bond (denoted as A) has  spin up,
while every pair of 
spins along a vertical nearest-neighbor bond (denoted as B) has spin down.

To construct a $T=0$ expansion about 
the Ising limit for this system,
one has to introduce an anisotropy parameter $\lambda$,
and write the Hamiltonian for the Heisenberg-Ising model as
\begin{equation}
H = H'_0 + \lambda V'~,  \label{Hising}
\end{equation}
where
\begin{eqnarray}
H'_0 &= &  \sum_{{\rm n.n.}} S_{i}^z S_{j}^z
+ x \sum_{{\rm n.n.n.}} S_{i}^z S_{j}^z +
 t [ \sum_{i\in B}  S_{i}^z - \sum_{i\in A}  S_{i}^z ] ~, \nonumber \\
V' &= & \sum_{{\rm n.n.}}
 ( S_{i}^x S_{j}^x + S_{i}^y S_{j}^y )  +
x \sum_{{\rm n.n.n.}} ( S_{i}^x S_{j}^x + S_{i}^y S_{j}^y )
- t  [ \sum_{i\in B}  S_{i}^z - \sum_{i\in A}  S_{i}^z ]~,
\end{eqnarray}
The last term
in both $H'_0$ and $V'$ is a local staggered field term, which can be
included to improve convergence.
The limits $\lambda =0$ and $\lambda =1$ correspond to the Ising model and
the isotropic Heisenberg model, respectively.
The operator $H'_0$ is taken as the unperturbed
Hamiltonian, with the unperturbed ground state being the
usual N\'{e}el state.
The operator $V'$ is treated as  a perturbation.
It flips a pair of spins on neighboring sites.

Ising series have been calculated for
the ground state energy per site, $E_0/N$ and
the staggered magnetization $M$,  
for several ratios of couplings $x$ and (simultaneously) for several values of $t$
 up to order $\lambda^{9}$.
The resulting series for $x=0.7, 0.8, 1, 2$ and $t=0$  are listed in
Table II; the series for other value of $x$ are
available upon request.

At the next stage of the analysis, we try to extrapolate
 the series to the isotropic point ($\lambda=1$)
for those values of the exchange coupling
parameters which lie within the N\'{e}el-ordered phase at $\lambda=1$.
For this purpose, we first transform the series to a new variable
\begin{equation}
\delta = 1- (1-\lambda)^{1/2}~,
\end{equation}
to remove the singularity at $\lambda=1$ predicted by the spin-wave theory. This
was first proposed by Huse\cite{hus} and was also
used in our earlier work on the square lattice case\cite{zhe1}.
We then use both integrated first-order inhomogeneous
differential approximants\cite{gut} and  Pad\'{e} approximants to
extrapolate the series to the isotropic point $\delta=1$  ($\lambda=1$).
The results of the Ising expansions will be presented in the next section.

\subsection{High-temperature series expansions}
We now turn to the finite-$T$ thermodynamic properties. We have developed
high-temperature series expansions $\beta=1/(k_B T)$ for the uniform magnetic
susceptibility $\chi (T)$
and the specific heat $C (T)$, for the system with arbitrary $x$,
\begin{eqnarray}
 T \chi (T) &=& {1 \over N} \sum_i \sum_j {{\rm Tr}
 S_i^z S_j^z e^{-\beta H } \over  {\rm Tr} e^{-\beta H } }~, \nonumber  \\ 
 &&  \\
 C (T) &=& {\partial U \over \partial T}~, \nonumber
\end{eqnarray}
where $N$ is the number of sites, and the internal
energy $U$ is defined by
\begin{equation}
U =  { {\rm Tr} H e^{-\beta H } \over  {\rm Tr} e^{-\beta H  } }~.
\end{equation}
The series for $\chi(T)$ and $C(T)$ were computed to order $\beta^{7}$ 
 for arbitrary $x$. The expansions have the following
general form:
\begin{equation}
\sum_{i=0}^{\infty} {\Big [} \sum_{j=0}^i c_{i,j} x^j {\Big ]} \beta^i~,
\end{equation}
where $c_{i,j}$ are the numerical coefficients.
The series coefficients are not presented here, 
but are available upon request.
Some of the coefficients can
be obtained by using the 
dimer expansion series at finite temperature discussed in the
next subsection. 
The high temperature series expansions were first carried out
by  Miyahara and Ueda\cite{miy98} up to order $\beta^{2}$.
Our results agree with these previous results,
and extend the series by 5 terms. 

\subsection{Dimer expansions at finite temperature}
The study of finite temperature properties via series expansions is
usually done by the high temperature expansion method as mentioned above,
where we expand in powers of $\beta$ for given ratio of exchange couplings.
This method often performs
poorly at low temperatures for many parameter regimes of interest.
To overcome this difficulty, one can develop
the dimer expansion at finite temperature, where we expand 
the thermodynamic quantities in power of $x$ for arbitrary
temperatures. This method has been used by one of us previously for the Hubbard model
\cite{otja92} and more recently also by Elstner and Singh\cite{rajiv98} for spin
models. This method
has shown excellent convergence for a wide range of coupling constants
at all temperatures for the bilayer Heisenberg model\cite{rajiv98} and for 
alternating spin-chains and spin ladders\cite{rajiv982}.

To get an expansion for thermodynamic quantities,
such as the susceptibility and the specific heat, in $x$ at arbitrary temperature
for the Hamiltonian in Eq.(\ref{Hdimer}), one basically needs to expand 
\begin{equation}
e^{-\beta (H_0 + x V)}
\end{equation}
in powers  of $x$.  This can be obtained by using the following relation
\begin{equation}
e^{-\beta (H_0 + x V)} = e^{-\beta H_0} \sum_n (-x)^n I_n~,
\end{equation}
where $I_n$ are the integrals given by
\begin{equation}
I_n = \int_0^{\beta} dt_1 \int_0^{t_1} dt_2 \cdots \int_0^{t_{n-1}} dt_n
\tilde{V} (t_1) \tilde{V} (t_2) \cdots \tilde{V}(t_n)~,
\end{equation}
with
\begin{equation}
\tilde{V} (t) = e^{t H_0} V e^{-t H_0}~.
\end{equation}
The integrations needed  in this expansion are of the type:
\begin{equation}
t^k e^{l t}~,
\end{equation}
and can  easily be integrated analytically.

The expansions for the susceptibility $\chi$ and the logarithm of the
partition function $\ln Z$ per site take the following general form\cite{rajiv98}:
\begin{eqnarray}
T \chi (x,\beta) &=& \sum_{n=0}  {f^{(\chi)}_n(\beta )\over 12 n!} \left( {x\over 12} \right)^n~, \nonumber \\
& & \label{chiZ} \\
\ln Z(x, \beta ) &=& 3\beta /4 + \ln Z_0 +  \sum_{n=1} 
{f^{(Z)}_n(\beta )\over 12 n!} \left( {x\over 12} \right)^n~, \nonumber
\end{eqnarray}
where $Z_0=1 + 3 \exp (- \beta)$, and the coefficient $f_n(\beta )$ for both $\chi$ and $\ln Z$
have the following general form
\begin{equation}
f_n (\beta) = \sum_{k=0}^n \sum_{l=0}^{n+1} c^{(n)}_{k,l} \beta^k Z_0^{-l}~, \label{finTc}
\end{equation}
where $c^{(n)}_{k,l}$ are expansion coefficients. The results up to order $n=6$ are
given in Table III.
From this expansion, one can recover the results of the high-temperature expansion 
and dimer expansions if 
we reexpand in powers of $\beta $, or $x$ at $T=0$. 

\section{Results}
Having obtained the series for the various expansions above we present in this section
the results of series analysis. We use 
integrated first-order inhomogeneous
differential approximants\cite{gut} and  Pad\'{e} approximants
to extrapolate the series. 

\subsection{Phase diagram}

The ground state energy per site $E_0/N$ is shown in Figure \ref{fig_e0}.
The  full points 
in the large $J'/J$ region are obtained from 
the Ising expansion, and the horizontal line corresponds to the eigenenergy
of the perfect dimer state (which is $E_0=3NJ/8$ exactly).
These curves cross at 
a transition point $x_{c}$, which corresponds to
a first order ground state phase transition, resulting from
a level crossing. The numerical estimate of $x_c=0.691(6)$ is 
a more precise estimate of the
 result $x_c=0.70(1)$ of Miyahara and Ueda\cite{miy98} 
discussed above. 

The staggered magnetization $M$ for those values of the exchange coupling
parameters which lie within the N\'{e}el-ordered phase at $\lambda=1$ is
shown in Fig. \ref{fig_M}. We can see that $M$ decreases as we turn on
$J$, and appears to vanish at around $J/J'\simeq 1.4$, 
near the transition point determined from the ground state energies.
The errors are too large to determine whether or not where is a finite discontinuity
at the transition.

\subsection{Triplet excitation spectrum}

From the dimer expansions, one can estimate  the triplet excitation spectrum
for those values of the exchange coupling
parameters which lie within the dimer phase.
The triplet excitation spectra for $x=0.6, 0.65$ and 0.678  are shown in Fig. \ref{fig_mk}.
We can see that the minimum energy gap is at $(k_x, k_y)=(0,0)$ (and
the equivalent point $(\pi,\pi)$). The bandwidth is quite small, indicating
that the triplet excitations are highly localized, but increases as $x_c$ is
approached.

Fig. \ref{fig_gap} shows the triplet excitation gap $\Delta =\Delta(0,0)$
 as a function of $x$. At $J'=0$, $\Delta/J$ is equal to 1 exactly, corresponding to a
single dimer excited to a triplet state. 
At the first order transition point $x_c$,
the gap is $\Delta = 0.14(5)$, where the uncertainty is largely associated
with the uncertainty of $x_c$.

\subsection{Thermodynamic properties}
Let us now discuss the thermodynamic properties.
Fig. \ref{fig_U} shows the internal energy per site $E/NJ$ vs.
temperature $k_BT/J$ obtained from the high temperature expansion and from
the dimer expansion at finite temperature for $x=0.5$. We can see that the
direct sum to the  dimer expansion series at finite temperature converges 
extremely well down to $T=0$, and it can recover the exact ground state
energy $E_0/NJ=-3/8$ at $T=0$. The results obtained from the integrated
differential approximants to the high-temperature series
expansion only converge well down to $k_BT/J \simeq 0.3$.

The results for the specific heat at $x=0.4$ obtained from the high-temperature expansion and
from the dimer expansion at finite temperature are shown at Fig. \ref{fig_C}.
We choose $x=0.4$ rather than $x=0.5$ 
because the specific heat series converges poorly for larger $x$.

Finally, the results for the susceptibility $\chi$ at $x=0.5$
are shown in Fig. \ref{fig_chi}.
Again, we can see that the results from a direct sum of the
 dimer expansion series at finite temperature
converges well down to $T=0$, (it converges less well near the peak position), while
the integrated differential approximants to the high-temperature
expansion series only converge to the position of the peak.

It is clear that the  dimer expansion at finite temperature gives much better
results than the  high-temperature expansion for those values of the exchange coupling
parameters which lie within the dimer phase. It is interesting to explore this also 
for ratios of the couplings lying within  the Ne\'el-ordered phase at $T=0$ 
($x>0.691$). In Fig. \ref{fig_U_1p5} and Fig. \ref{fig_chi_1p5}, we present
the results for internal energy and susceptibility for $x=1.5$, well outside
the dimer phase at $T=0$, and one can see that dimer expansions at finite
temperature still give  good convergence in the high temperature region.
Evidently at high temperature the system is highly disordered and 
both methods include contributions from all states.

\subsection{Compound  SrCu$_2$(BO$_3$)$_2$}
Finally we compare the experimental results for the compound SrCu$_2$(BO$_3$)$_2$  with the
theoretical calculations to get an estimate of the exchange constants.
Kageyama {\it et al.}\cite{kag98} determined the excitation gap $\Delta = 25$ K
from the NMR relaxation rate. They also  measure the temperature
dependence of the magnetic susceptibility, and this implies a slightly smaller gap:
$\Delta =(19\pm 1)$ K.
Since there is some uncertainty in the energy gap, 
to determine the exchange constants we begin with the magnetic susceptibility.
To convert the theoretical $\chi(T)$ into units emu/Cu mol in experiment,  we
multiply our $\chi$ (calculated with $J=1$) by
$N_A g^2 \mu_B^2/k_B$, with $\mu_B$ the Bohr magneton,
$k_B$ the Boltzmann constant, $N_A$ Avogadro's number,
and $g$ the effective $g$-factor.
There are three fitting parameters: $J$ and $x$ (or $J'$) and $g$.
Our goal is try to find a proper parameter set $(J,x,g)$ which gives the minimum value of
\begin{equation}
P = \sum_{T_i} [\chi^{\rm exp}(T_i) - \chi^{\rm theo}(T_i) ]^2~,
\end{equation}
where $\chi^{\rm exp}$, $\chi^{\rm theo}(T_i)$ are the
experimental and theoretical susceptibilities, respectively, and the summation
is over all experimental points $T_i$.
As the theoretical susceptibility from the 6th order dimer expansion at finite temperature
converges well above temperatures corresponding to a physical temperature $T_0\sim 100$ K,
we restrict the minimisation function to temperatures $T>T_0$. We then scan the
values of $P$ around the expected parameter region for SrCu$_2$(BO$_3$)$_2$ to
locate the minimum $P_{\rm min}=1.14\times 10^{-9} ({\rm emu/Cu Mol})^2$
at $J=92.0$ K, $x=0.563$ and $g=2.104$. 
This is shown as the full point in Fig. \ref{fig_fit_exp}

Since there is some error in both the experimental and theoretical $\chi$,
we cannot expect that the true parameters for SrCu$_2$(BO$_3$)$_2$ are located exactly 
at the minimum point of $P$, (in fact, the location of the minimum point depends
on the chosen  value of $T_0$), we need to consider the energy gap also.
In the 3-dimensional parameter space $(J,x,g)$, there is a long tube-like
region which has $P\leq 5 P_{\rm min}$. The intersection with the plane $g=2.108$
is shown as the dashed lines  in Fig. \ref{fig_fit_exp}(a).
If we fix the value of $J$ and vary $x$ and $g$, 
we obtain curves 
$x(J)$ and $g(J)$ which give near minimum values of
$P$. These curves $x(J)$ and $g(J)$ and the corresponding 
$P$ are given in Fig.  \ref{fig_fit_exp}. We can see that along
$x(J)$, $P$ only changes slightly by about 10\%,  
and $g$ is about $2.11$ almost independent of $J$.
To further 
determine the parameters for SrCu$_2$(BO$_3$)$_2$, we need to consider the energy gap 
(which relates to the low-temperature behavior of $\chi(T)$).
From Fig. \ref{fig_gap}, we can get two curves in $(J,x)$ space where 
the energy gap is 19 K and 25 K, respectively,  as 
shown in Fig. \ref{fig_fit_exp}. We see that they cross with the optimal
$x(J)$ curve obtained above at
$$
J=82.0 {\rm K}, \quad x=0.678, \quad g=2.108
$$ 
if the energy gap is 19 K, or at
$$
J=83.2 {\rm K},\quad  x=0.664, \quad g=2.108
$$
if the energy gap is 25 K.
We believe this should be the best fit to both
$\chi(T)$ and the energy gap. 
 
If we take the  energy gap to be 19 K,
the comparison of the experimental data with theoretical
calculations is shown in Fig. \ref{fig_exp1} and Fig. \ref{fig_exp2}.
We can see the fit is extremely good in the high temperature region.
We note that the dimer expansion sums tend to show sharp peaks,
which are not visible in either the experimental data or the high-temperature
expansions. We suspect that these are an artefact, as the results will
become more sensitive to the truncation of the dimer expansion at finite
order near the transition point in $x$.
The ratio of couplings $x$ obtained here is similar to that obtained
by Miyahara and Ueda\cite{miy98}, but they found 
$J=100$ K which is nearly 20\% larger than our estimate.
They obtained their results by considering the paramagnetic 
Curie-Weiss constant $\theta$.
The Curie-Weiss constant given by the high-temperature expansions is $\theta = (J+4J')/4$.
From experimental data, the susceptibility at high temperature
can be fitted with a Curie-Weiss constant $\theta=92.5$ K and an effective 
$g$-factor $g=2.14$.\cite{kag98} The result curve for $(J+4J')/4=92.5$ K 
is shown as a dotted line in Fig. 11(a). 
The discrepancy between the two estimates is 
perhaps due to uncertainty in extracting the 
Curie-Weiss constant from experimental data at finite 
temperature. Our estimates imply the Curie-Weiss 
constant to be $\theta =76.1$ K.

\section{Conclusions}
Several different series expansions have been calculated to high order for this model:
high-temperature expansions, an Ising expansion at zero temperature,
and dimer expansion at both zero and finite temperature.
The first-order transition from the dimer phase to a N\'eel-ordered phase has
been found to occur at $J'/J=0.691(6)$, in good 
agreement with the original estimate 0.70(1) of
Miyahara and Ueda\cite{miy98}. The ground-state energy shows a sharp
and distinct break in slope at that point, 
indicative of a first-order transition,
and the triplet spin gap undergoes 
a small but definite discontinuity: the triplet spin gap is $\Delta =0.14(5)$
at the transition point.
A discontinuity in the N\'eel phase magnetization is less certain, 
but is not excluded by our results.

The model has been fitted to the experimental susceptibility data for SrCu$_2$(BO$_3$)$_2$\cite{kag98}, 
with parameters
$g=2.108$, $J=82.0$ K and $x=0.678$ if the energy gap is 19 K or
$g=2.108$, $J=83.2$ K and $x=0.664$ if the energy gap is 25 K. 
A detailed and accurate fit can be obtained at
high temperature, and a reasonable fit at low temperatures is also obtained.
The fitted value of $x=(J'/J)$ is about 0.67,
which is indeed very close to the transition point 0.691.

We should summarize what is new in this work. The Ising expansion
has been developed for this model for the first time, and this allows a
more accurate determination of the phase transition point. We have
also estimated the spontaneous magnetization in the Ne\'el phase.
The finite temperature dimer expansion has not been computed
previously. This provides the most reliable theoretical
estimate of the high temperature susceptibility and is used to estimate
the experimental parameters of the real system. Apart from this
we have also substantially increased the length of the dimer
series at $T=0$ and the conventional high temperature expansion.
The former has allowed us to present the first calculation of the
full triplet excitation spectrum for this system.

\acknowledgments
This work forms part of a research project supported by a grant 
from the Australian Research Council. 
The computation has been performed on Silicon Graphics Power 
Challenge and Convex machines, and we thank the New South Wales 
Centre for Parallel Computing for facilities and assistance
with the calculations. We would like to thank Prof.  Ueda and
Dr. Kageyama for providing the experimental data on SrCu$_2$(BO$_3$)$_2$.


\begin{figure}[htb]
\caption{(a) Lattice structure of the Cu$^{2+}$ spins of 
SrCu$_2$(BO$_3$)$_2$. The nearest-neighbor bonds
are expressed by solid lines and the next-nearest-neighbor
bonds by dashed lines. (b) Elementary unit for interaction 
between a pair of nearest-neighbor bonds. The dotted lines
denote the additional coupling
($J_2$) which still allow the perfect dimer state as eigenstate.
}
\label{fig_lat}
\end{figure}

\begin{figure}[htb]
\caption{The ground-state energy per 
site $E_0/NJ$  as 
function of $J'/J$.
The solid horizontal line is the energy of the perfect dimer state, while
the solid points with error bars are the estimates from
the  Ising expansion.
}
\label{fig_e0}
\end{figure}

\begin{figure}[htb]
\caption{The staggered magnetization $M$ 
{\it versus} $J/J'$.
The solid points with error bars are the estimates from
the Ising expansions.
}
\label{fig_M}
\end{figure}

\begin{figure}[htb]
\caption{Plot of 
triplet excitation  spectrum 
$\Delta(k_x, k_y)$ 
(derived from the dimer expansions) along high-symmetry
cuts through the Brillouin zone for  coupling ratios
$x=0.6, 0.65, 0.678$.}
\label{fig_mk}
\end{figure}

\begin{figure}[htb]
\caption{The triplet excitation  gap
$\Delta = \Delta(0, 0)$ as a function of coupling ratios
$x$ derived from the dimer expansion.
Several different integrated differential approximants to
the series are shown. The errorbar indicates the gap at 
the critical point.}
\label{fig_gap}
\end{figure}

\begin{figure}[htb]
\caption{Internal energy per site $E/NJ$ vs. temperature $k_B T/J$ for $x=1/2$.
The solid and dashed lines are the direct sum of the dimer expansion series at 
finite temperature to orders from 2 to 6, while
the dotted lines are several different integrated differential approximants to
the high-temperature series.}
\label{fig_U}
\end{figure}

\begin{figure}[htb]
\caption{Specific heat per site $C$ vs. temperature $k_BT/J$ for $x=0.4$.
The solid and dashed lines are the direct sum of the dimer expansion series at 
finite temperature to orders from 2 to 6, while
the dotted lines are several different integrated differential approximants to
the high-temperature series..}
\label{fig_C}
\end{figure}

\begin{figure}[htb]
\caption{Susceptibility $\chi$ per site $\chi$ vs. temperature $k_BT/J$ for $x=1/2$.
The solid and dashed lines are direct sum of the dimer expansion series at 
finite temperature to orders from 2 to 6, while
the dotted lines are several different integrated differential approximants to
the high-temperature series..}
\label{fig_chi}
\end{figure}

\begin{figure}[htb]
\caption{Internal energy per site $E/NJ$ vs. temperature $k_B T/J$ for $x=3/2$.
The solid and dashed lines are direct sum of series of dimer expansion at 
finite temperature to order from 2 to 6, while
the dotted lines are several different integrated differential approximants to
high-temperature series.}
\label{fig_U_1p5}
\end{figure}

\begin{figure}[htb]
\caption{Susceptibility $\chi$ per site $\chi$ vs. temperature $k_BT/J$ for $x=3/2$.
The solid and dashed lines are the direct sum of the dimer expansion series at 
finite temperature to orders from 2 to 6, while
the dotted lines are several different integrated differential approximants to
the high-temperature series..}
\label{fig_chi_1p5}
\end{figure}

\begin{figure}[htb]
\caption{The two nearly horizontal solid lines in the lower window 
indicate  where 
the energy gap is 19 K and 25 K, respectively. The other 
solid lines in both upper and lower windows
indicate for a given value of $J$, the values of $(x,g)$ which give a minimum 
value of $P$, and the corresponding value of $P$ is also given in
the upper window. The region bounded by two dashed lines in the lower window
indicates the range in parameter space $(J,x,g=2.108)$ that $P$ is less than or
equal to $5 P_{\rm min}$, where $P_{\rm min}$ is the minimum of $P$ in the whole
parameter space, indicated  by the solid box point in the lower window.
The dotted line in the lower window indicates where the Weiss constant
$\theta = (J+4 J')/4 =92.5$ K.
}
\label{fig_fit_exp}
\end{figure}

\begin{figure}[htb]
\caption{Comparison of the calculated temperature dependence of the susceptibility 
with experimental data\protect\cite{kag98} (open points). 
The solid and dashed lines are direct sums of the dimer expansion series at 
finite temperature to orders from 2 to 6 for Heisenberg model with
parameters $J=82$ K, $x=0.678$ and $g=2.108$.}
\label{fig_exp1}
\end{figure}

\begin{figure}[htb]
\caption{Same as Fig. \ref{fig_exp1}, but the experimental data are compared with
the estimates from high-temperature expansions (the solid lines representing
various integrated differential approximants to the high-temperature series.)}
\label{fig_exp2}
\end{figure}

\widetext

\setdec 0.00000000000
\begin{table}
\squeezetable
\caption{Series coefficients for the dimer expansion of the triplet
 excitation spectrum $\Delta (k_x, k_y) =$ 
$ J \sum_{k,n,m} a_{k,n,m} x^{k} \cos (m k_x + n k_y) $.
 Nonzero coefficients $a_{k,n,m}$
up to order $k=15$  are listed.}\label{tabisigap}
\begin{tabular}{rr|rr|rr|rr}
\multicolumn{1}{c}{(k,n,m)} &\multicolumn{1}{c|}{$a_{k,n,m}$} 
&\multicolumn{1}{c}{(k,n,m)} &\multicolumn{1}{c|}{$a_{k,n,m}$}
& \multicolumn{1}{c}{(k,n,m)} &\multicolumn{1}{c|}{$a_{k,n,m}$}
& \multicolumn{1}{c}{(k,n,m)} &\multicolumn{1}{c}{$a_{k,n,m}$} \\
\hline
 ( 0, 0, 0) &\dec   1.000000000  &( 7, 1, 1) &\dec $-$3.819444444$\times 10^{-2}$ &(11, 1,-1) &\dec $-$6.345273723$\times 10^{-2}$ &(13, 1, 2) &\dec   1.039117707$\times 10^{-3}$ \\
 ( 2, 0, 0) &\dec $-$1.000000000  &( 8, 1, 1) &\dec $-$3.602430556$\times 10^{-2}$ &(12, 1,-1) &\dec $-$1.767150368$\times 10^{-1}$ &(14, 1, 2) &\dec   2.816852119$\times 10^{-3}$ \\
 ( 3, 0, 0) &\dec $-$5.000000000$\times 10^{-1}$ &( 9, 1, 1) &\dec $-$7.375458140$\times 10^{-3}$ &(13, 1,-1) &\dec $-$2.169670151$\times 10^{-1}$ &(15, 1, 2) &\dec   6.075705255$\times 10^{-3}$ \\
 ( 4, 0, 0) &\dec $-$1.250000000$\times 10^{-1}$ &(10, 1, 1) &\dec $-$2.159589603$\times 10^{-4}$ &(14, 1,-1) &\dec $-$9.902031964$\times 10^{-2}$ &(12, 2, 2) &\dec   5.358886719$\times 10^{-4}$ \\
 ( 5, 0, 0) &\dec   1.562500000$\times 10^{-1}$ &(11, 1, 1) &\dec $-$6.345273723$\times 10^{-2}$ &(15, 1,-1) &\dec   3.702751072$\times 10^{-2}$ &(13, 2, 2) &\dec   1.970669593$\times 10^{-3}$ \\
 ( 6, 0, 0) &\dec   2.343750000$\times 10^{-2}$ &(12, 1, 1) &\dec $-$1.767150368$\times 10^{-1}$ &(10, 2,-1) &\dec $-$7.866753472$\times 10^{-5}$ &(14, 2, 2) &\dec   3.433763327$\times 10^{-3}$ \\
 ( 7, 0, 0) &\dec $-$3.687065972$\times 10^{-1}$ &(13, 1, 1) &\dec $-$2.169670151$\times 10^{-1}$ &(11, 2,-1) &\dec $-$1.746396665$\times 10^{-4}$ &(15, 2, 2) &\dec   2.148257717$\times 10^{-3}$ \\
 ( 8, 0, 0) &\dec $-$6.348922164$\times 10^{-1}$ &(14, 1, 1) &\dec $-$9.902031964$\times 10^{-2}$ &(12, 2,-1) &\dec $-$3.807362686$\times 10^{-4}$ &(10, 1,-2) &\dec $-$7.866753472$\times 10^{-5}$ \\
 ( 9, 0, 0) &\dec $-$3.904155213$\times 10^{-1}$ &(15, 1, 1) &\dec   3.702751072$\times 10^{-2}$ &(13, 2,-1) &\dec $-$1.039117707$\times 10^{-3}$ &(11, 1,-2) &\dec $-$1.746396665$\times 10^{-4}$ \\
 (10, 0, 0) &\dec   1.468893169$\times 10^{-1}$ &(10, 2, 1) &\dec   7.866753472$\times 10^{-5}$ &(14, 2,-1) &\dec $-$2.816852119$\times 10^{-3}$ &(12, 1,-2) &\dec $-$3.807362686$\times 10^{-4}$ \\
 (11, 0, 0) &\dec   2.410994431$\times 10^{-1}$ &(11, 2, 1) &\dec   1.746396665$\times 10^{-4}$ &(15, 2,-1) &\dec $-$6.075705255$\times 10^{-3}$ &(13, 1,-2) &\dec $-$1.039117707$\times 10^{-3}$ \\
 (12, 0, 0) &\dec $-$6.102249664$\times 10^{-1}$ &(12, 2, 1) &\dec   3.807362686$\times 10^{-4}$ &(14, 3,-1) &\dec $-$4.682591353$\times 10^{-7}$ &(14, 1,-2) &\dec $-$2.816852119$\times 10^{-3}$ \\
 (13, 0, 0) &\dec $-$1.761368461  &(13, 2, 1) &\dec   1.039117707$\times 10^{-3}$ &(15, 3,-1) &\dec $-$5.703426062$\times 10^{-7}$ &(15, 1,-2) &\dec $-$6.075705255$\times 10^{-3}$ \\
 (14, 0, 0) &\dec $-$1.770029604  &(14, 2, 1) &\dec   2.816852119$\times 10^{-3}$ &(10, 0, 2) &\dec $-$8.680555556$\times 10^{-4}$ &(12, 2,-2) &\dec   5.358886719$\times 10^{-4}$ \\
 (15, 0, 0) &\dec $-$1.691084766$\times 10^{-1}$ &(15, 2, 1) &\dec   6.075705255$\times 10^{-3}$ &(11, 0, 2) &\dec $-$2.002495660$\times 10^{-3}$ &(13, 2,-2) &\dec   1.970669593$\times 10^{-3}$ \\
 (10, 2, 0) &\dec $-$2.486617477$\times 10^{-4}$ &(14, 3, 1) &\dec $-$4.682591353$\times 10^{-7}$ &(12, 0, 2) &\dec $-$7.904745973$\times 10^{-4}$ &(14, 2,-2) &\dec   3.433763327$\times 10^{-3}$ \\
 (11, 2, 0) &\dec $-$3.280526620$\times 10^{-4}$ &(15, 3, 1) &\dec $-$5.703426062$\times 10^{-7}$ &(13, 0, 2) &\dec   6.061904493$\times 10^{-3}$ &(15, 2,-2) &\dec   2.148257717$\times 10^{-3}$ \\
 (12, 2, 0) &\dec   9.612269774$\times 10^{-4}$ &( 6, 1,-1) &\dec $-$2.083333333$\times 10^{-2}$ &(14, 0, 2) &\dec   1.576995064$\times 10^{-2}$ &(14, 1, 3) &\dec $-$9.042245370$\times 10^{-6}$ \\
 (13, 2, 0) &\dec   4.323950651$\times 10^{-3}$ &( 7, 1,-1) &\dec $-$3.819444444$\times 10^{-2}$ &(15, 0, 2) &\dec   1.850923153$\times 10^{-2}$ &(15, 1, 3) &\dec $-$2.671619241$\times 10^{-5}$ \\
 (14, 2, 0) &\dec   6.550286294$\times 10^{-3}$ &( 8, 1,-1) &\dec $-$3.602430556$\times 10^{-2}$ &(10, 1, 2) &\dec   7.866753472$\times 10^{-5}$ &(14, 1,-3) &\dec $-$9.042245370$\times 10^{-6}$ \\
 (15, 2, 0) &\dec   8.172418539$\times 10^{-4}$ &( 9, 1,-1) &\dec $-$7.375458140$\times 10^{-3}$ &(11, 1, 2) &\dec   1.746396665$\times 10^{-4}$ &(15, 1,-3) &\dec $-$2.671619241$\times 10^{-5}$ \\
 ( 6, 1, 1) &\dec $-$2.083333333$\times 10^{-2}$ &(10, 1,-1) &\dec $-$2.159589603$\times 10^{-4}$ &(12, 1, 2) &\dec   3.807362686$\times 10^{-4}$ &\\
\end{tabular}
\end{table}

\begin{table}
\squeezetable
\setdec 0.0000000000000
\caption{Series coefficients  of the 
Ising expansion for  the
ground-state energy per site $E_0/(NJ)$ and the staggered magnetization $M$,
 for $J'/J=0.7,0.8,1,2$ 
and $t=0$. Nonzero coefficients of $\lambda^i$
up to order $i=9$  are listed.}
 \label{tab2}
\begin{tabular}{r|rrrr}
\multicolumn{1}{c|}{$i$} 
&\multicolumn{1}{c}{$J'/J=0.7$}&\multicolumn{1}{c}{$J'/J=0.8$} 
&\multicolumn{1}{c}{$J'/J=1$} &\multicolumn{1}{c}{$J'/J=2$}  \\
\tableline
\multicolumn{1}{c|}{}&\multicolumn{4}{c}{$E_0/(JN)$}\\
  0 &\dec $-$2.2500000000$\times 10^{-1}$ &\dec $-$2.7500000000$\times 10^{-1}$ &\dec $-$3.7500000000$\times 10^{-1}$ &\dec $-$8.7500000000$\times 10^{-1}$ \\
  1 &\dec  0.0000000000  &\dec  0.0000000000  &\dec  0.0000000000  &\dec  0.0000000000  \\
  2 &\dec $-$2.2272727273$\times 10^{-1}$ &\dec $-$2.2857142857$\times 10^{-1}$ &\dec $-$2.5000000000$\times 10^{-1}$ &\dec $-$4.0000000000$\times 10^{-1}$ \\
  3 &\dec  1.0123966942$\times 10^{-1}$ &\dec  8.1632653061$\times 10^{-2}$ &\dec  6.2500000000$\times 10^{-2}$ &\dec  4.0000000000$\times 10^{-2}$ \\
  4 &\dec $-$1.3065004904$\times 10^{-1}$ &\dec $-$7.4186951609$\times 10^{-2}$ &\dec $-$3.5416666667$\times 10^{-2}$ &\dec $-$6.1067821068$\times 10^{-3}$ \\
  5 &\dec  2.6804197345$\times 10^{-1}$ &\dec  1.2604414182$\times 10^{-1}$ &\dec  5.0052083333$\times 10^{-2}$ &\dec  1.0890524707$\times 10^{-2}$ \\
  6 &\dec $-$5.3530295140$\times 10^{-1}$ &\dec $-$1.9791154855$\times 10^{-1}$ &\dec $-$6.1429832176$\times 10^{-2}$ &\dec $-$1.2407230180$\times 10^{-2}$ \\
  7 &\dec  1.1672953875  &\dec  3.2781449128$\times 10^{-1}$ &\dec  7.4708752894$\times 10^{-2}$ &\dec  8.8405316048$\times 10^{-3}$ \\
  8 &\dec $-$2.9587106052  &\dec $-$6.5298672657$\times 10^{-1}$ &\dec $-$1.1569637564$\times 10^{-1}$ &\dec $-$1.0969498862$\times 10^{-2}$ \\
  9 &\dec  7.7545598044  &\dec  1.3549175235  &\dec  1.8737141791$\times 10^{-1}$ &\dec  1.0540011561$\times 10^{-2}$ \\
\tableline
\multicolumn{1}{c|}{}&\multicolumn{4}{c}{Magnetization $M$} \\
  0 &\dec  5.0000000000$\times 10^{-1}$ &\dec  5.0000000000$\times 10^{-1}$ &\dec  5.0000000000$\times 10^{-1}$ &\dec  5.0000000000$\times 10^{-1}$ \\
  1 &\dec  0.0000000000  &\dec  0.0000000000  &\dec  0.0000000000  &\dec  0.0000000000  \\
  2 &\dec $-$4.0495867769$\times 10^{-1}$ &\dec $-$3.2653061224$\times 10^{-1}$ &\dec $-$2.5000000000$\times 10^{-1}$ &\dec $-$1.6000000000$\times 10^{-1}$ \\
  3 &\dec  3.6814425244$\times 10^{-1}$ &\dec  2.3323615160$\times 10^{-1}$ &\dec  1.2500000000$\times 10^{-1}$ &\dec  3.2000000000$\times 10^{-2}$ \\
  4 &\dec $-$1.3553013187  &\dec $-$6.2496725406$\times 10^{-1}$ &\dec $-$2.4361111111$\times 10^{-1}$ &\dec $-$5.1163207419$\times 10^{-2}$ \\
  5 &\dec  3.4305010340  &\dec  1.2470397038  &\dec  3.5369444444$\times 10^{-1}$ &\dec  3.4797358726$\times 10^{-2}$ \\
  6 &\dec $-$9.1962657747  &\dec $-$2.5644449411  &\dec $-$5.5283436053$\times 10^{-1}$ &\dec $-$4.6108122016$\times 10^{-2}$ \\
  7 &\dec  2.5590774140$\times 10^{1}$ &\dec  5.4148695622  &\dec  8.5367555708$\times 10^{-1}$ &\dec  4.0349371222$\times 10^{-2}$ \\
  8 &\dec $-$7.8821305450$\times 10^{1}$ &\dec $-$1.3000427014$\times 10^{1}$ &\dec $-$1.5762664485  &\dec $-$5.9375271517$\times 10^{-2}$ \\
  9 &\dec  2.4067476616$\times 10^{2}$ &\dec  3.1003113436$\times 10^{1}$ &\dec  2.8719797054  &\dec  6.2701777926$\times 10^{-2}$ \\
\end{tabular}
\end{table}

\begin{table}
\squeezetable
\caption{Series coefficients for the dimer expansion at finite temperature of the magnetic
susceptibility $\chi$ and the logarithm of the partition function
$\ln Z$. Nonzero coefficients $c^{(n)}_{k,l}$ in
Eq. (\ref{finTc})
up to order $n=6$  are listed.}\label{tabfinT}
\begin{tabular}{rr|rr|rr|rr}
\multicolumn{1}{c}{(n,k,l)} &\multicolumn{1}{c|}{$c^{(n)}_{k,l}$}
&\multicolumn{1}{c}{(n,k,l)} &\multicolumn{1}{c|}{$c^{(n)}_{k,l}$}
&\multicolumn{1}{c}{(n,k,l)} &\multicolumn{1}{c|}{$c^{(n)}_{k,l}$}
&\multicolumn{1}{c}{(n,k,l)} &\multicolumn{1}{c}{$c^{(n)}_{k,l}$} \\
\hline
\multicolumn{8}{c}{$T \chi(x,\beta)$} \\
 ( 0, 0, 0) &              4  &( 4, 3, 1) &      $-$20689920  &( 5, 2, 3) &    $-$6519859200  &( 6, 5, 2) &   126045573120  \\
 ( 0, 0, 1) &             $-$4  &( 4, 4, 1) &       $-$3239424  &( 5, 3, 3) &    28138690560  &( 6, 6, 2) &     5370983424  \\
 ( 1, 1, 0) &            $-$64  &( 4, 1, 2) &       23417856  &( 5, 4, 3) &     1845657600  &( 6, 1, 3) &  $-$386640691200  \\
 ( 1, 1, 1) &            128  &( 4, 2, 2) &       50377728  &( 5, 5, 3) &      $-$56417280  &( 6, 2, 3) & $-$2967429427200  \\
 ( 1, 1, 2) &            $-$64  &( 4, 3, 2) &       56881152  &( 5, 1, 4) &    $-$1577410560  &( 6, 3, 3) & $-$1905248194560  \\
 ( 2, 1, 0) &            640  &( 4, 4, 2) &        1916160  &( 5, 2, 4) &     6189373440  &( 6, 4, 3) & $-$2606694359040  \\
 ( 2, 2, 0) &           1344  &( 4, 1, 3) &      $-$22579200  &( 5, 3, 4) &   $-$36834969600  &( 6, 5, 3) &     9843886080  \\
 ( 2, 1, 1) &          $-$4224  &( 4, 2, 3) &     $-$109221888  &( 5, 4, 4) &     1241210880  &( 6, 6, 3) &    $-$5610921984  \\
 ( 2, 2, 1) &          $-$3456  &( 4, 3, 3) &      $-$68871168  &( 5, 5, 4) &       33269760  &( 6, 1, 4) &   412034826240  \\
 ( 2, 1, 2) &           7680  &( 4, 4, 3) &        1858560  &( 5, 1, 5) &      188743680  &( 6, 2, 4) &  3178633052160  \\
 ( 2, 2, 2) &           2880  &( 4, 1, 4) &        7593984  &( 5, 2, 5) &    $-$2117468160  &( 6, 3, 4) &  4207696773120  \\
 ( 2, 1, 3) &          $-$4096  &( 4, 2, 4) &      105916416  &( 5, 3, 5) &    23898193920  &( 6, 4, 4) &  3593735331840  \\
 ( 2, 2, 3) &           $-$768  &( 4, 3, 4) &       37969920  &( 5, 4, 5) &    $-$2151383040  &( 6, 5, 4) &  $-$331372339200  \\
 ( 3, 1, 0) &          16128  &( 4, 4, 4) &       $-$2433024  &( 5, 5, 5) &       36126720  &( 6, 6, 4) &    15752116224  \\
 ( 3, 2, 0) &         $-$39168  &( 4, 2, 5) &      $-$37355520  &( 5, 3, 6) &    $-$6102712320  &( 6, 1, 5) &  $-$160650362880  \\
 ( 3, 3, 0) &         $-$36864  &( 4, 3, 5) &       $-$7667712  &( 5, 4, 6) &      747110400  &( 6, 2, 5) & $-$1772148326400  \\
 ( 3, 1, 1) &        $-$103680  &( 4, 4, 5) &         712704  &( 5, 5, 6) &      $-$21626880  &( 6, 3, 5) & $-$5437322035200  \\
 ( 3, 2, 1) &         359424  &( 5, 1, 0) &     $-$190402560  &( 6, 1, 0) &    $-$3039897600  &( 6, 4, 5) & $-$2693322178560  \\
 ( 3, 3, 1) &          99072  &( 5, 2, 0) &      $-$75018240  &( 6, 2, 0) &    35239587840  &( 6, 5, 5) &   441996410880  \\
 ( 3, 1, 2) &         179712  &( 5, 3, 0) &     $-$157562880  &( 6, 3, 0) &    24536862720  &( 6, 6, 5) &   $-$22042312704  \\
 ( 3, 2, 2) &        $-$933120  &( 5, 4, 0) &     $-$142878720  &( 6, 4, 0) &    17376768000  &( 6, 1, 6) &     9059696640  \\
 ( 3, 3, 2) &         $-$78336  &( 5, 5, 0) &      $-$43966464  &( 6, 5, 0) &     8583413760  &( 6, 2, 6) &   402165596160  \\
 ( 3, 1, 3) &         $-$92160  &( 5, 1, 1) &     1436820480  &( 6, 6, 0) &     1929157632  &( 6, 3, 6) &  3619867852800  \\
 ( 3, 2, 3) &         944640  &( 5, 2, 1) &     $-$283760640  &( 6, 1, 1) &    $-$2405099520  &( 6, 4, 6) &  1007563898880  \\
 ( 3, 3, 3) &           6912  &( 5, 3, 1) &     2067609600  &( 6, 2, 1) &  $-$386275184640  &( 6, 5, 6) &  $-$240553820160  \\
 ( 3, 2, 4) &        $-$331776  &( 5, 4, 1) &     1148129280  &( 6, 3, 1) &  $-$162191831040  &( 6, 6, 6) &    13220904960  \\
 ( 3, 3, 4) &           9216  &( 5, 5, 1) &      119927808  &( 6, 4, 1) &  $-$204870942720  &( 6, 3, 7) &  $-$958440407040  \\
 ( 4, 1, 0) &        1290240  &( 5, 1, 2) &    $-$3521018880  &( 6, 5, 1) &   $-$63168215040  &( 6, 4, 7) &  $-$140236554240  \\
 ( 4, 2, 0) &        1256448  &( 5, 2, 2) &     2806732800  &( 6, 6, 1) &    $-$5717993472  &( 6, 5, 7) &    48625090560  \\
 ( 4, 3, 0) &        2377728  &( 5, 3, 2) &   $-$11009249280  &( 6, 1, 2) &   131641528320  &( 6, 6, 7) &    $-$2901934080  \\
 ( 4, 4, 0) &        1185024  &( 5, 4, 2) &    $-$2687846400  &( 6, 2, 2) &  1509814702080  &  \\
 ( 4, 1, 1) &       $-$9722880  &( 5, 5, 2) &      $-$67313664  &( 6, 3, 2) &   611100979200  &  \\
 ( 4, 2, 1) &      $-$10973184  &( 5, 1, 3) &     3663267840  &( 6, 4, 2) &  1026448035840  &  \\
\hline
\multicolumn{8}{c}{$\ln Z(x,\beta)$} \\ 
 ( 2, 1, 0) &           $-$576  &( 4, 2, 2) &      $-$21150720  &( 5, 5, 2) &      $-$16692480  &( 6, 4, 2) &   151663104000  \\
 ( 2, 2, 0) &            576  &( 4, 3, 2) &       10824192  &( 5, 1, 3) &     $-$313528320  &( 6, 5, 2) &   $-$29455488000  \\
 ( 2, 1, 1) &           2880  &( 4, 4, 2) &       $-$1534464  &( 5, 2, 3) &     1816888320  &( 6, 6, 2) &     2645913600  \\
 ( 2, 2, 1) &          $-$1152  &( 4, 1, 3) &        2322432  &( 5, 3, 3) &       58475520  &( 6, 1, 3) &   143954288640  \\
 ( 2, 1, 2) &          $-$2304  &( 4, 2, 3) &       22975488  &( 5, 4, 3) &     $-$145981440  &( 6, 2, 3) &   262998005760  \\
 ( 2, 2, 2) &            576  &( 4, 3, 3) &      $-$12441600  &( 5, 5, 3) &       15068160  &( 6, 3, 3) &   676278097920  \\
 ( 3, 1, 0) &         $-$10368  &( 4, 4, 3) &        1465344  &( 5, 2, 4) &     $-$663552000  &( 6, 4, 3) &  $-$444080517120  \\
 ( 3, 2, 0) &         $-$10368  &( 4, 2, 4) &       $-$9289728  &( 5, 4, 4) &       51425280  &( 6, 5, 3) &   100333209600  \\
 ( 3, 3, 0) &           1728  &( 4, 3, 4) &        4644864  &( 5, 5, 4) &       $-$4700160  &( 6, 6, 3) &    $-$8033458176  \\
 ( 3, 1, 1) &          51840  &( 4, 4, 4) &        $-$470016  &( 6, 1, 0) &    $-$6057815040  &( 6, 1, 4) &   $-$53031075840  \\
 ( 3, 2, 1) &          20736  &( 5, 1, 0) &       87713280  &( 6, 2, 0) &    $-$5008988160  &( 6, 2, 4) &  $-$219410104320  \\
 ( 3, 3, 1) &          $-$3456  &( 5, 2, 0) &      $-$42716160  &( 6, 3, 0) &    $-$8885790720  &( 6, 3, 4) &  $-$814218117120  \\
 ( 3, 1, 2) &         $-$41472  &( 5, 3, 0) &      $-$54743040  &( 6, 4, 0) &    $-$1159557120  &( 6, 4, 4) &   620258549760  \\
 ( 3, 2, 2) &         $-$10368  &( 5, 4, 0) &       $-$2073600  &( 6, 5, 0) &      187038720  &( 6, 5, 4) &  $-$143287418880  \\
 ( 3, 3, 2) &           1728  &( 5, 5, 0) &        $-$656640  &( 6, 6, 0) &      $-$37324800  &( 6, 6, 4) &    10433193984  \\
 ( 4, 1, 0) &        $-$705024  &( 5, 1, 1) &     $-$516948480  &( 6, 1, 1) &    49705436160  &( 6, 2, 5) &    66886041600  \\
 ( 4, 2, 0) &       $-$1575936  &( 5, 2, 1) &      664796160  &( 6, 2, 1) &    38078760960  &( 6, 3, 5) &   537954877440  \\
 ( 4, 3, 0) &        $-$124416  &( 5, 3, 1) &      167961600  &( 6, 3, 1) &    90482780160  &( 6, 4, 5) &  $-$419908976640  \\
 ( 4, 4, 0) &         $-$13824  &( 5, 4, 1) &      $-$38983680  &( 6, 4, 1) &   $-$17612328960  &( 6, 5, 5) &    94038589440  \\
 ( 4, 1, 1) &        4105728  &( 5, 5, 1) &        6981120  &( 6, 5, 1) &     1647267840  &( 6, 6, 5) &    $-$6303744000  \\
 ( 4, 2, 1) &        9040896  &( 5, 1, 2) &      742763520  &( 6, 6, 1) &     $-$164395008  &( 6, 3, 6) &  $-$147786301440  \\
 ( 4, 3, 1) &       $-$2903040  &( 5, 2, 2) &    $-$1775416320  &( 6, 1, 2) &  $-$134570833920  &( 6, 4, 6) &   110839726080  \\
 ( 4, 4, 1) &         552960  &( 5, 3, 2) &     $-$171694080  &( 6, 2, 2) &  $-$143543715840  &( 6, 5, 6) &   $-$23463198720  \\
 ( 4, 1, 2) &       $-$5723136  &( 5, 4, 2) &      135613440  &( 6, 3, 2) &  $-$333825546240  &( 6, 6, 6) &     1459814400  \\
\end{tabular}
\end{table}


\begin{references}
\bibitem[*]{byline1}e-mail address: w.zheng@unsw.edu.au
\bibitem[\dag]{byline2}e-mail address: c.hamer@unsw.edu.au
\bibitem[\ddag]{byline3}e-mail address: otja@newt.phys.unsw.edu.au

\bibitem{srcu2o3}M. Azuma {\it et al.}, Phys. Rev. Let. {\bf 73},
3643(1994).
\bibitem{cav2o5}H. Iwase, {\it et al.}, J. Phys. Soc. Jpn {\bf 65},
2397(1996).
\bibitem{vopo}A.W. Garret {\it et al.}, Phys. Rev. Lett. 
{\bf 79}, 745(1997).

\bibitem{cuHpCl}G. Chaboussant {\it et al.}, Phys. Rev. B {\bf 55},
3046(1997); Phys. Rev. Lett. {\bf 79}, 925(1997).


\bibitem{cav4o9}S. Taniguchi, {\it et al.}, J. Phys. Soc. Jpn, {\bf 64},
2758(1995).

\bibitem{kag98}H. Kageyama, {\it et al.}, submitted to PRL.

\bibitem{miy98}S. Miyahara and K. Ueda, cond-mat/9807075.

\bibitem{shastry} B.S. Shastry and B. Sutherland,
Physica 108{\bf B}, 1069 (1981).

\bibitem{he90}H.X. He, C.J. Hamer and J. Oitmaa, J. Phys. A {\bf 23}, 1775(1990).
\bibitem{gel90}M. P. Gelfand, R.R.P. Singh, and D.A, Huse, J. of Stat. Phys. {\bf 59},
1093(1990).
\bibitem{gelmk}M. P. Gelfand, Solid State Commun. {\bf 98}, 11(1996).

\bibitem{otja92}J.A. Henderson, J. Oitmaa and M.C. Ashley,
Phys. Rev. B {\bf 46}, 6328(1992).

\bibitem{rajiv98}N. Elstner and R.R.P. Singh, Phys. Rev. B {\bf 57}, 7740(1998).

\bibitem{hus}D.A. Huse, Phys. Rev. B{\bf 37}, 2380(1988).
\bibitem{zhe1}W.H. Zheng, J. Oitmaa and C.J. Hamer, Phys. Rev. B {\bf 43},
8321(1991).
\bibitem{gut}A.J. Guttmann, in ``Phase Transitions and Critical
Phenomena'', Vol. 13 ed. C. Domb and J. Lebowitz (New York, Academic, 1989).

\bibitem{rajiv982}N. Elstner and R.R.P. Singh, cond-mat/9803085.

\end{references}
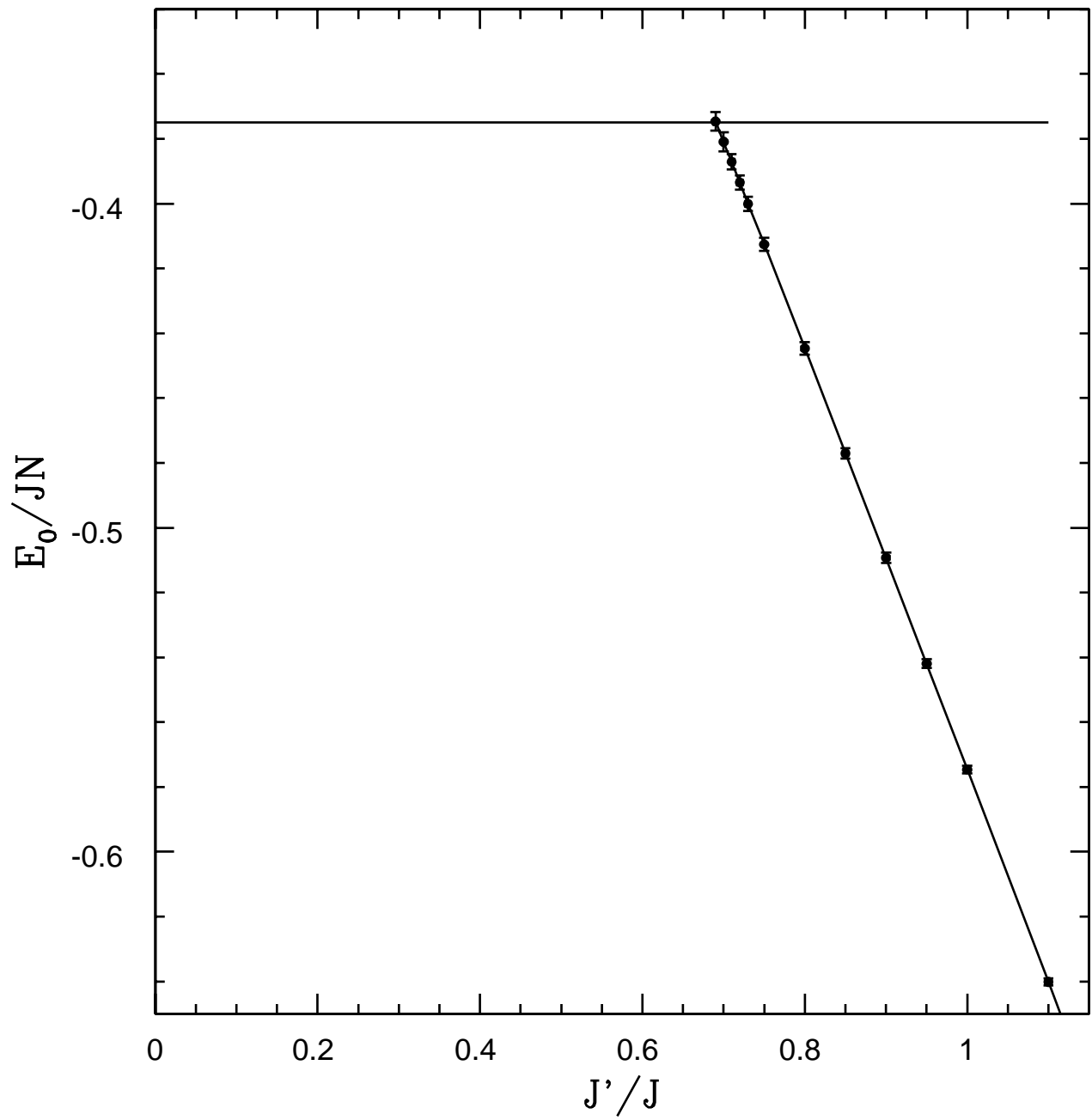
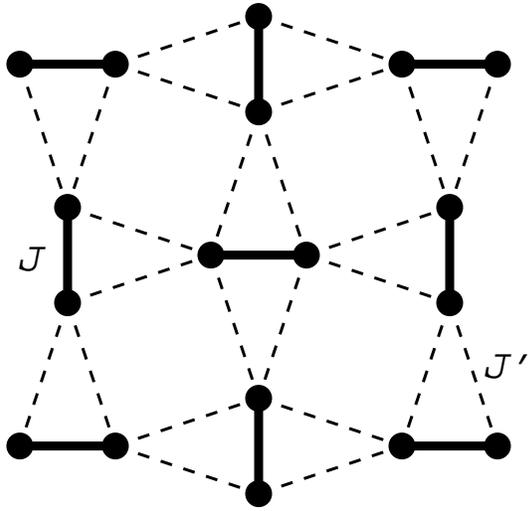
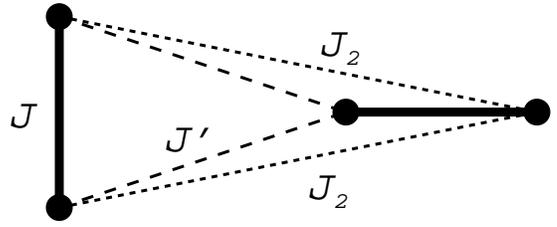
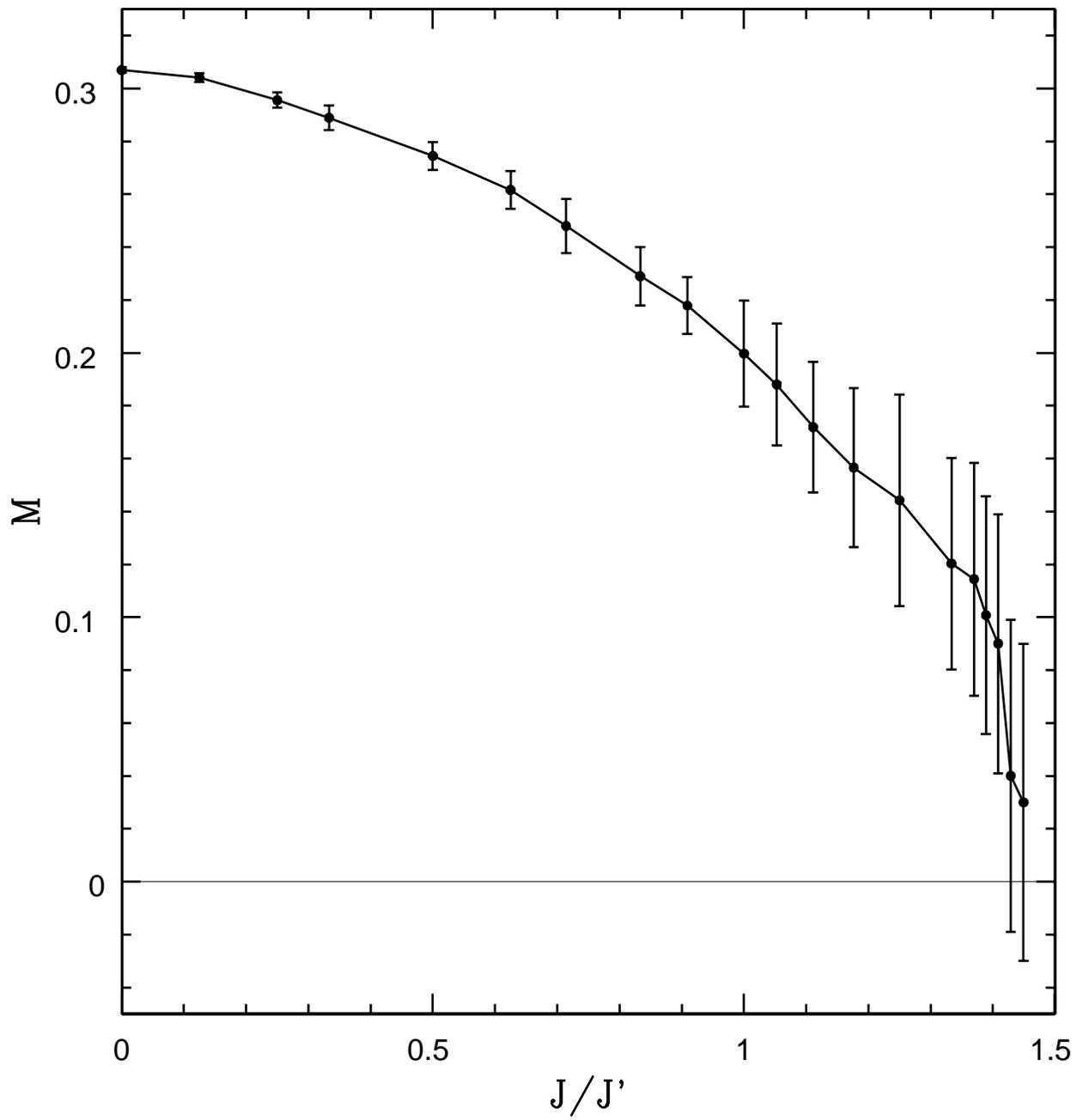
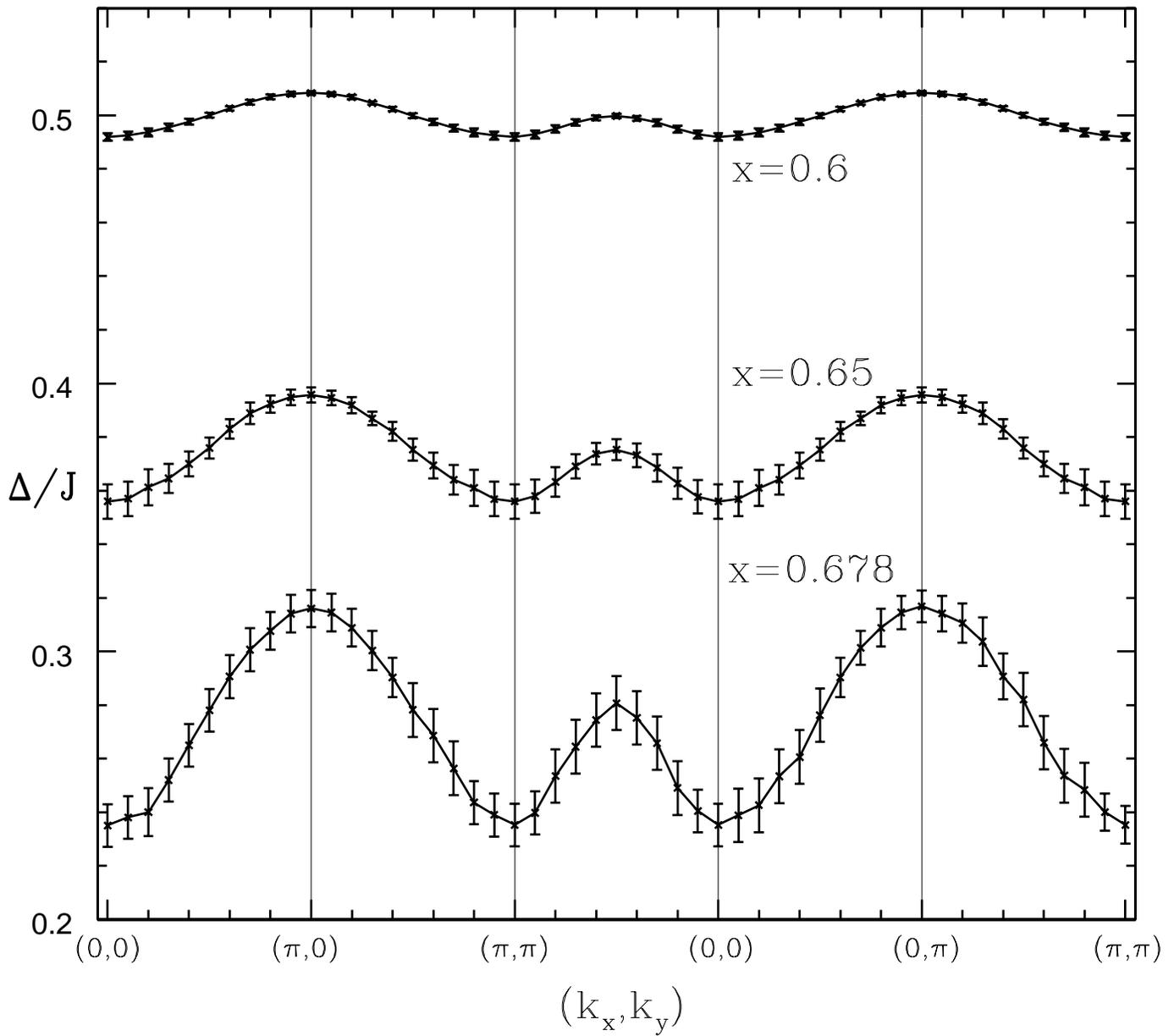
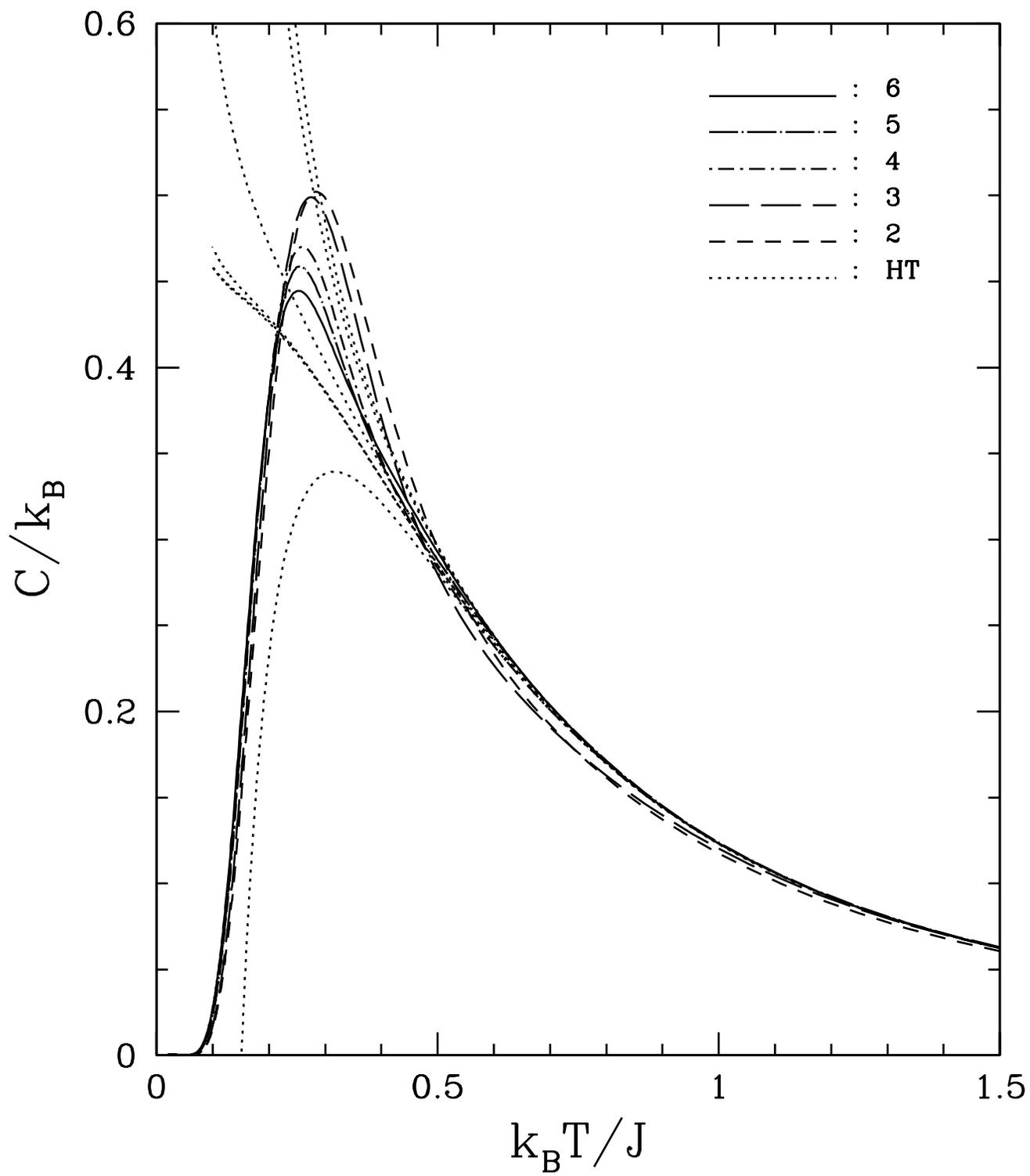
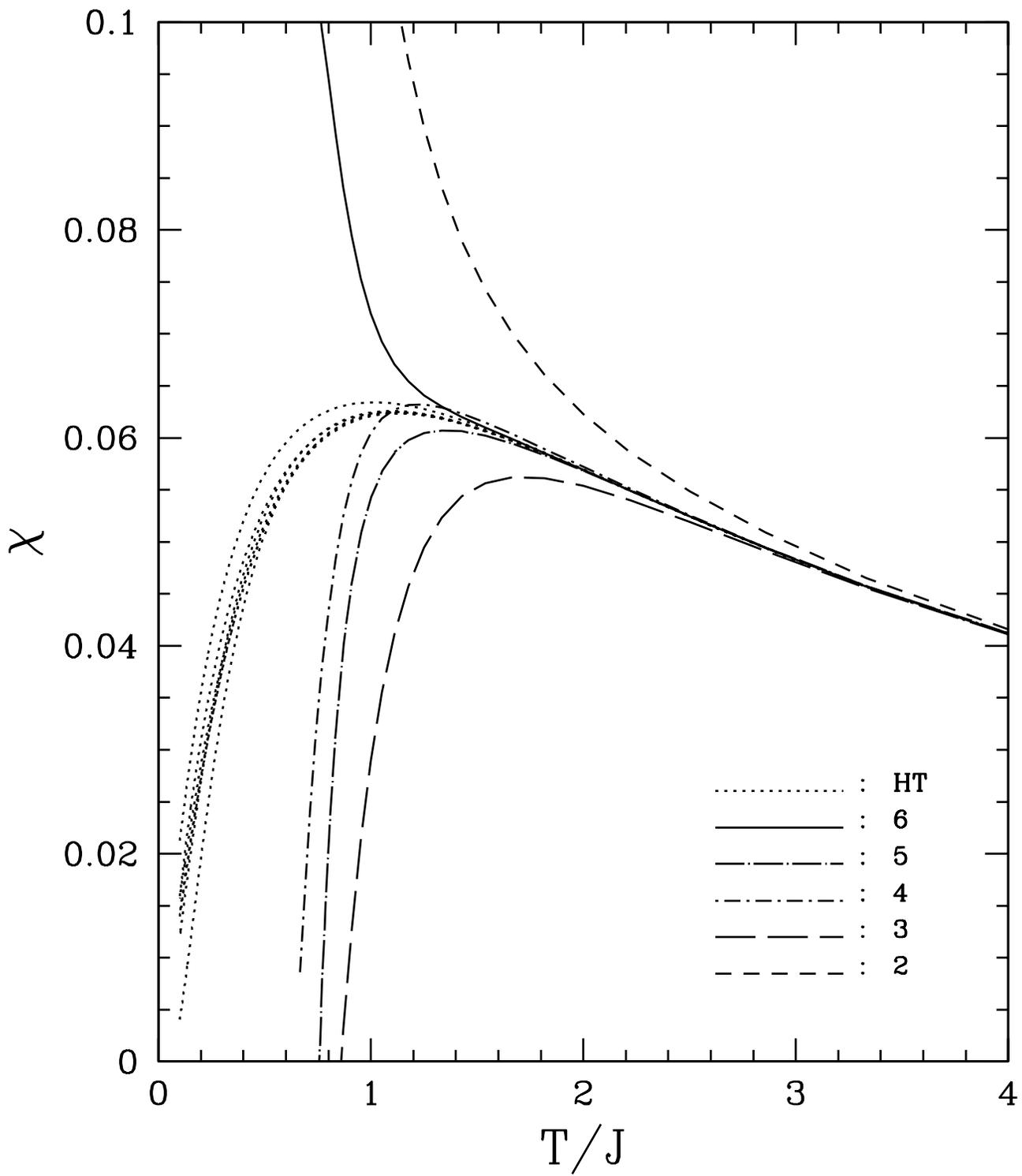
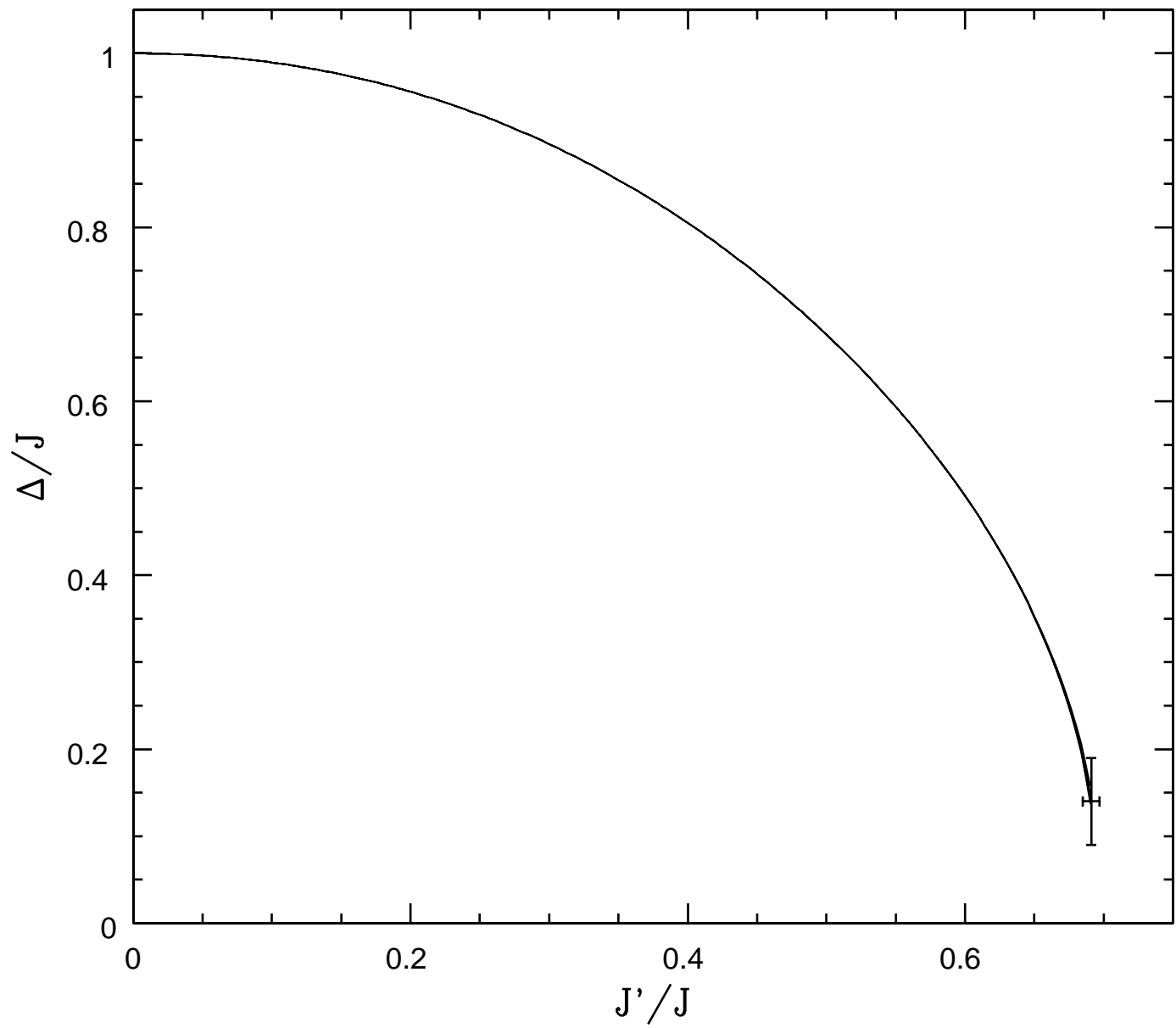
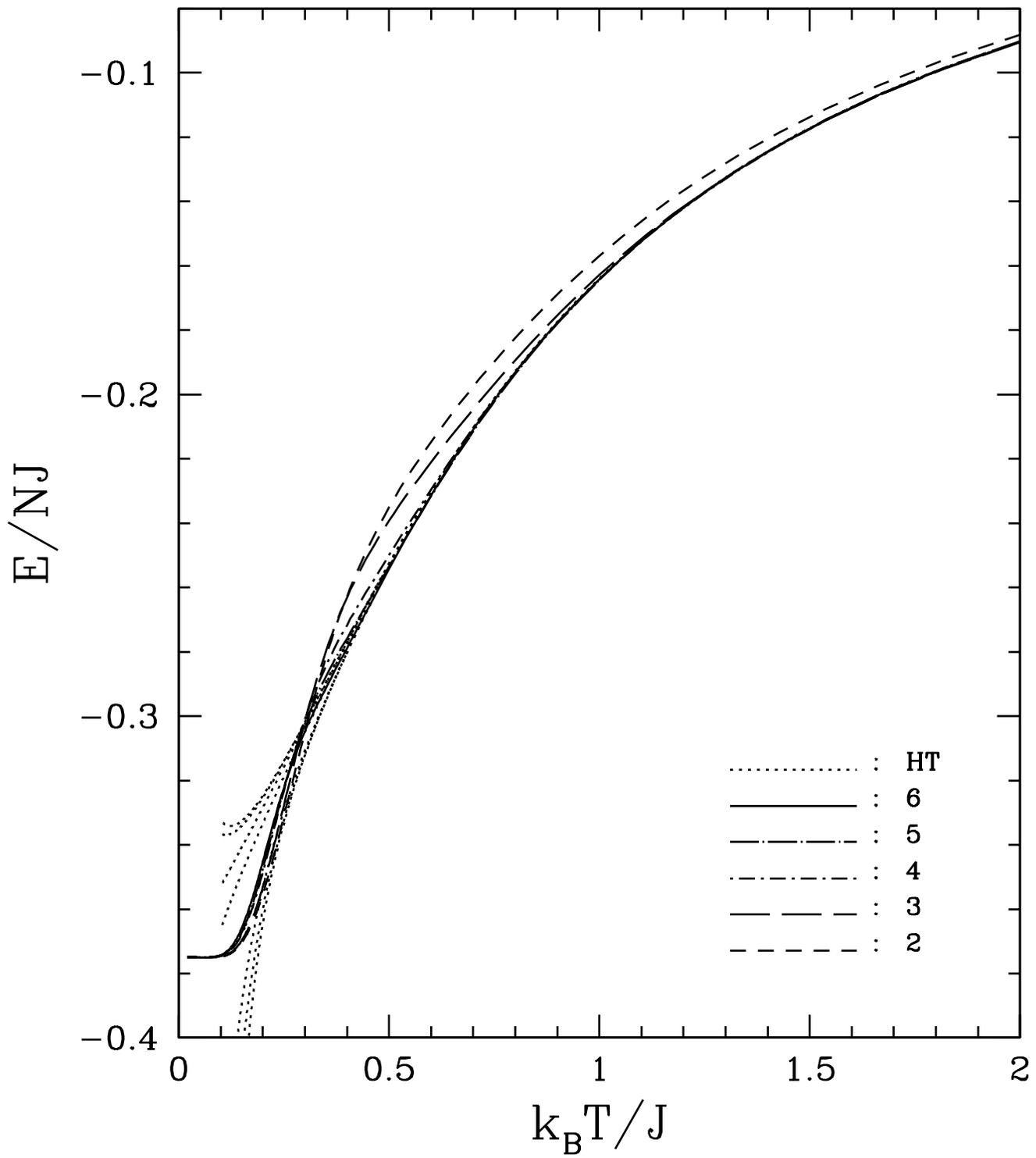
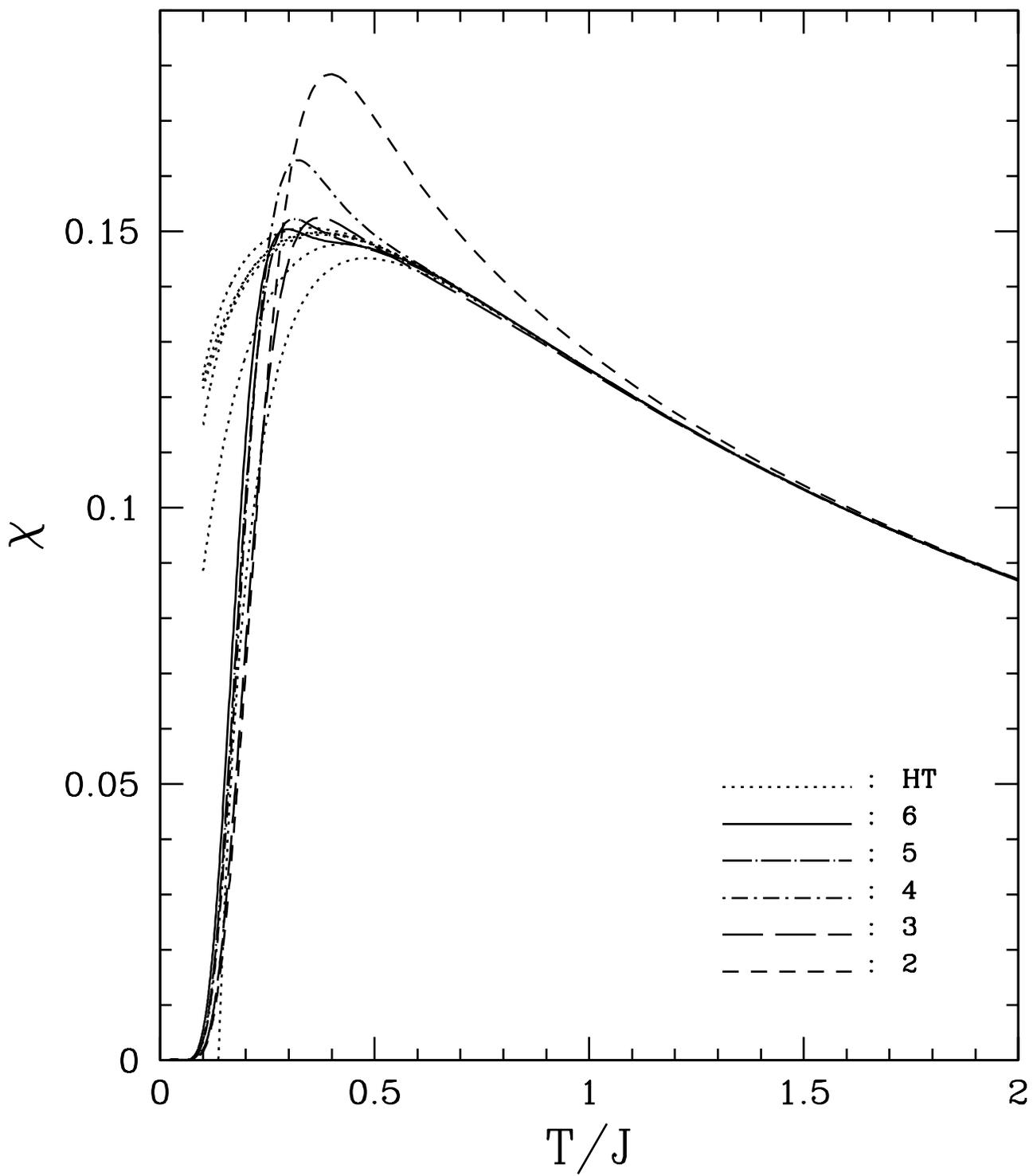
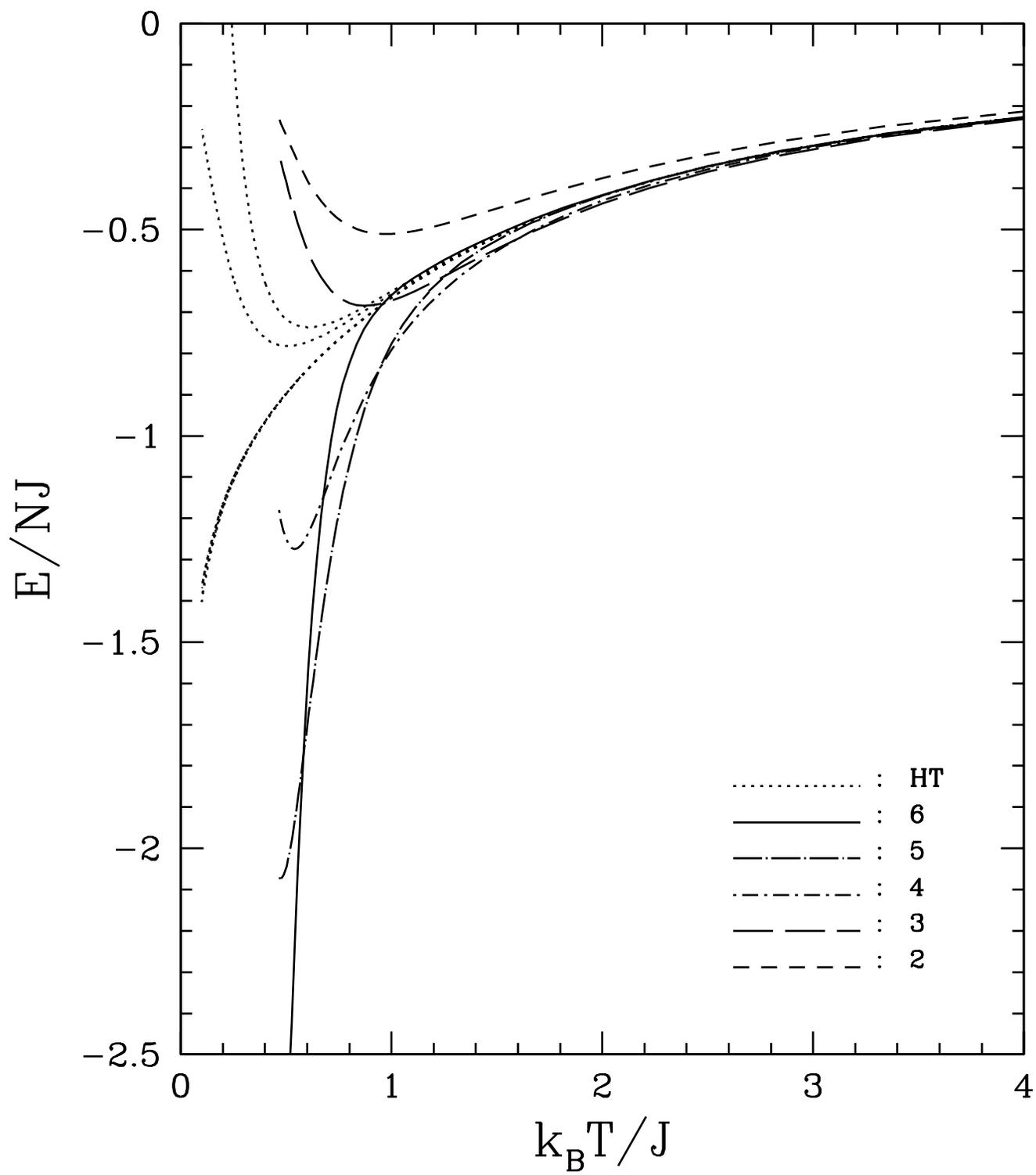
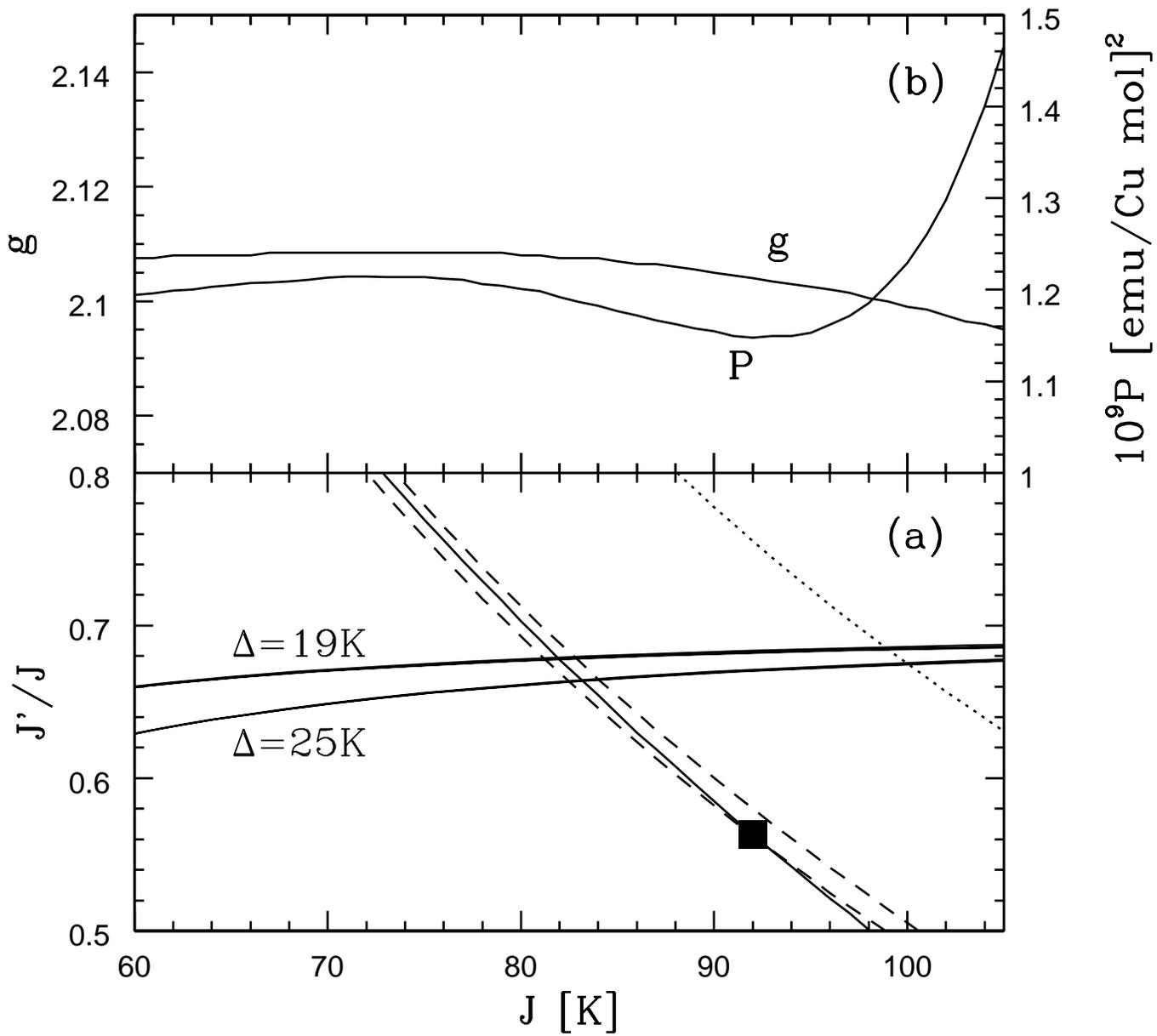
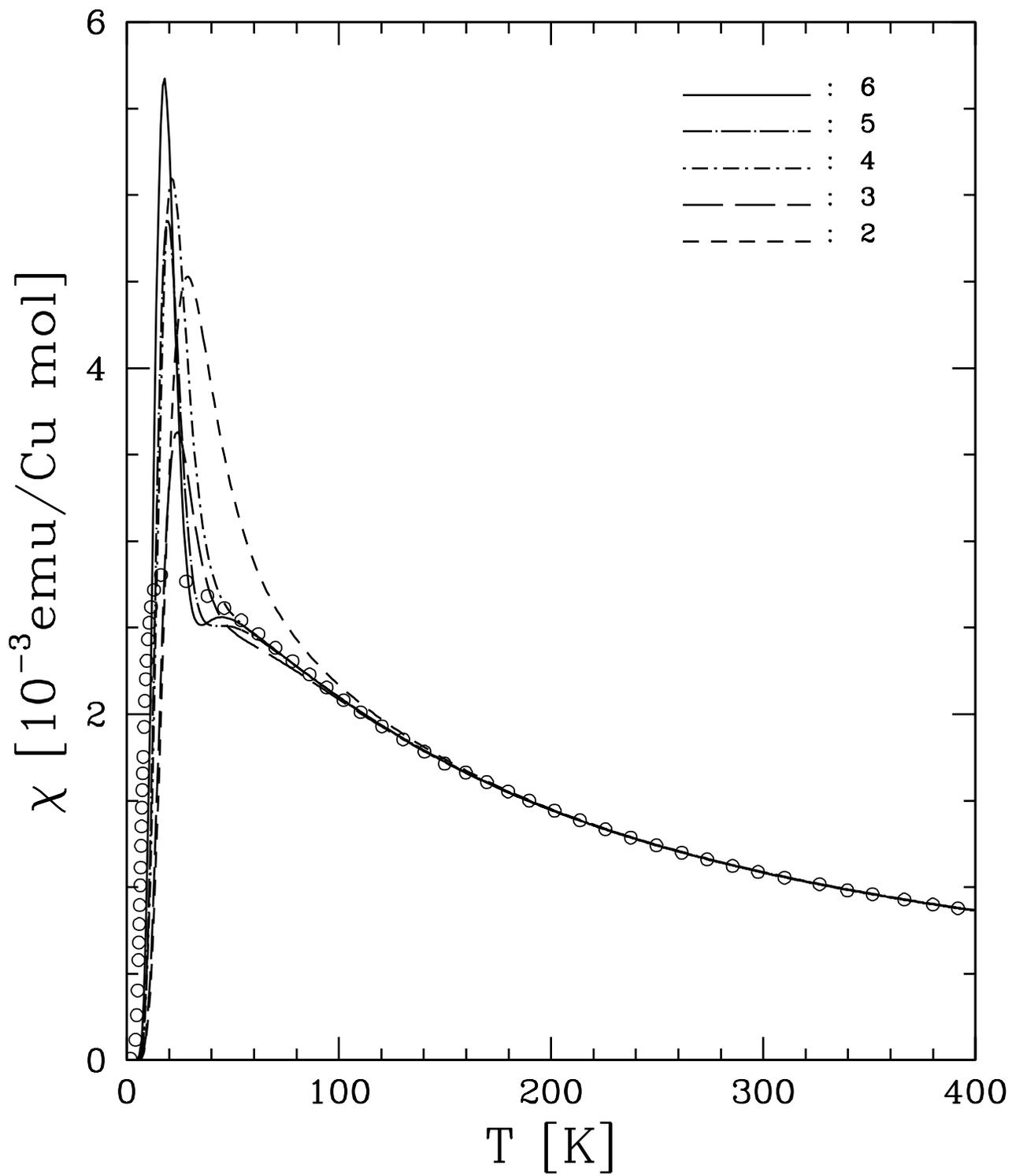
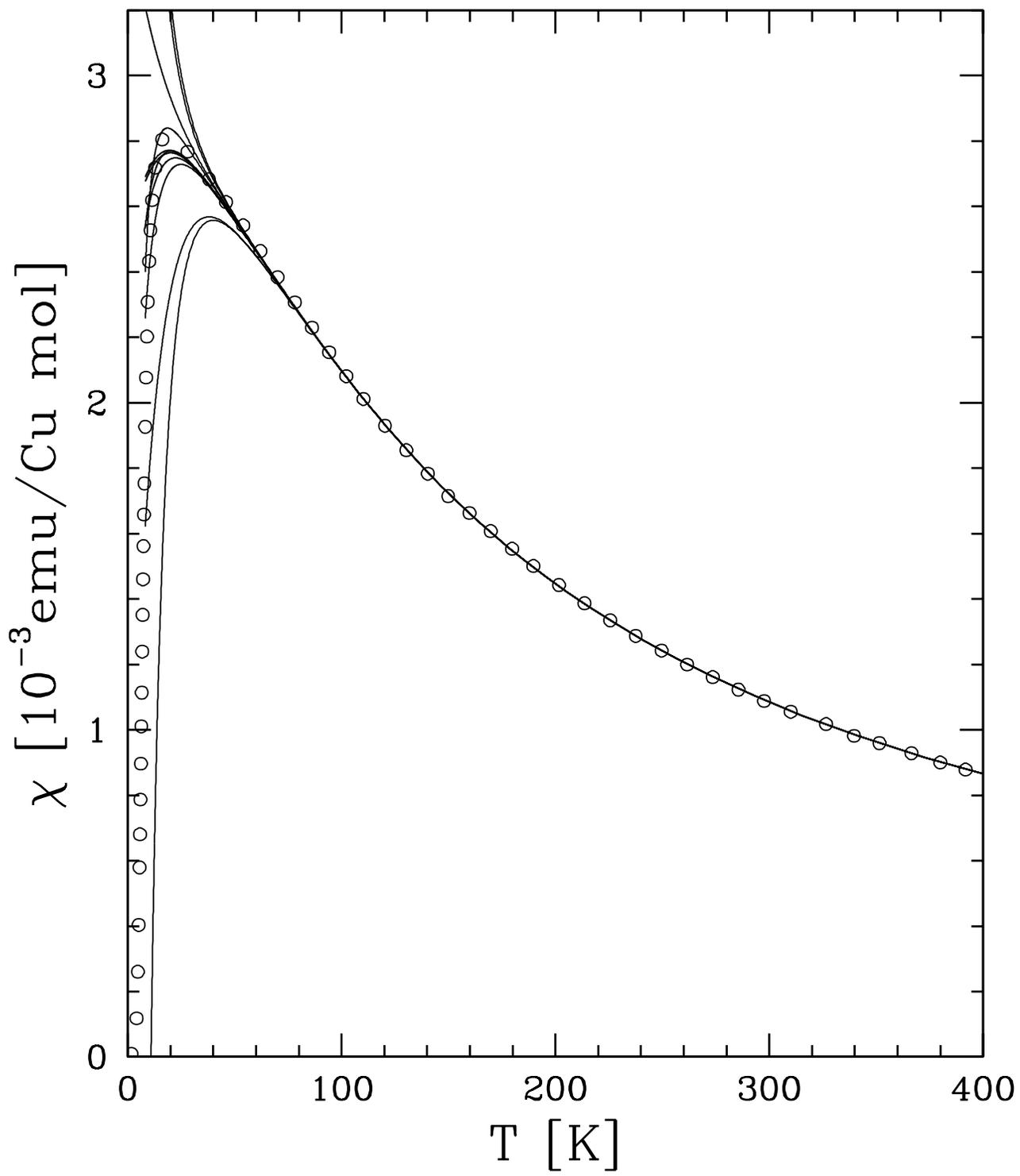
\end{document}